# Predicted Diurnal Variation of the Deuterium to Hydrogen Ratio in Water at the Surface of Mars Caused by Mass Exchange with the Regolith

Renyu Hu[1,2]

[1] Jet Propulsion Laboratory, California Institute of Technology, Pasadena, CA 91109, USA, renyu.hu@jpl.nasa.gov

[2] Division of Planetary and Geological Science, California Institute of Technology, Pasadena, CA 91125, USA

Address:
4800 Oak Grove Dr.
Jet Propulsion Laboratory, MS 169-237
Pasadena, CA 91109, USA
renyu.hu@jpl.nasa.gov

**Highlights**

- We predict diurnal variation of the D/H at the surface of Mars of ~1000‰.
- The variation is mainly driven by adsorption and desorption of regolith particles.
- Stable isotopes can be used to pinpoint regolith-atmosphere exchange on Mars.

**Abstract**

Regolith on Mars exchanges water with the atmosphere on a diurnal basis and this process causes significant variation in the abundance of water vapor at the surface. While previous studies of regolith-atmosphere exchange focus on the abundance, recent in-situ experiments and remote sensing observations measure the isotopic composition of the atmospheric water. We are therefore motivated to investigate isotopic water exchange between the atmosphere and the regolith and determine its effect on the deuterium to hydrogen ratio (D/H) of the atmosphere. We model transport of water in the regolith and regolith-atmosphere exchange by solving a transport equation including regolith adsorption, condensation, and diffusion. The model calculates equilibrium fractionation between HDO and $H_2O$ in each of these processes. The fractionation in adsorption is caused by the difference in the latent heat of adsorption, and that of condensation is caused by the difference in the vapor pressure. Together with a simple, bulk-aerodynamic boundary layer model, we simulate the diurnal variation of the D/H near the planetary surface. We find that the D/H can vary by 300 – 1400‰ diurnally in the equatorial and mid-latitude locations, and the magnitude is greater at a colder location or season. The variability is mainly driven by adsorption and desorption of regolith particles, and its diurnal trend features a drop in the early morning, a rise to the peak value during the daytime, and a second drop in the late afternoon and evening, tracing the water vapor flow into and out from the regolith. The predicted D/H variation can be tested with in-situ measurements. As such, our calculations suggest stable isotope analysis to be a powerful tool in pinpointing regolith-atmosphere exchange of water on Mars.

# 1. Introduction

The regolith is a main reservoir of volatiles on Mars. This reservoir stores substantial amount of water by physical adsorption (Fanale & Cannon 1971) and as ice at high latitudes (Mellon & Jakosky 1993). Due to low surface temperatures, the Martian regolith has great volatile adsorbing and storing capacities. The Gamma-Ray Spectrometer (GRS) suite onboard the Mars Odyssey mission has detected high water content poleward of 60° latitude in both hemispheres (Boynton et al. 2002). The Dynamic Albedo of Neutrons (DAN) instrument onboard the Curiosity rover has determined that the water content of the top ~1m regolith to be 1~5 wt. % (Mitrofanov et al. 2014).

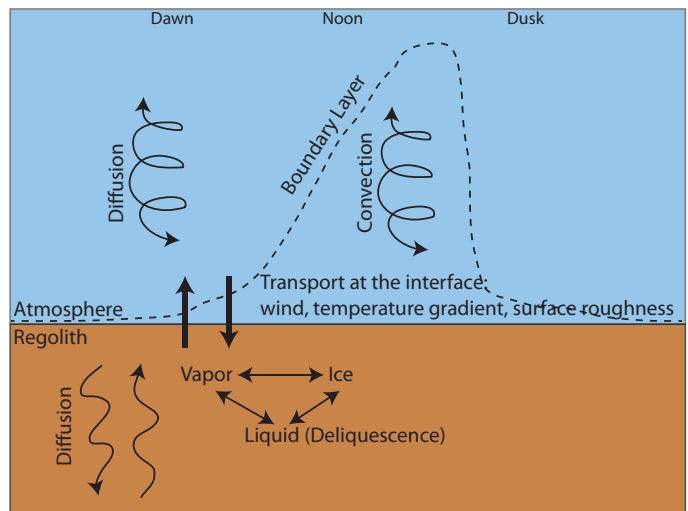

The adsorbed water is released to the atmosphere and re-adsorbed to the regolith on a diurnal basis (e.g., Flasar & Goody 1976; Zent et al. 1993; Savijärvi 1995; also see Fig. 1). The diurnal cycle is mainly driven by large variations of the surface temperature, and that the adsorption coefficient is a strong function of the temperature.

Tremendous progress has been made to understand regolith-atmosphere exchange of water on Mars by measuring the variation of near-surface humidity (e.g., Jakosky et al. 1997; Zent et al. 2016; Savijärvi et al. 2016). Both Phoenix and Curiosity rovers are equipped with relative humidity sensors: Thermal and Electrical Conductivity Probe (TECP) on Phoenix (Zent et al. 2010); Rover Environmental Monitoring Station (REMS) on Curiosity (Gómez-Elvira et al. 2012). While the relative humidity is primarily controlled by the ambient temperature (e.g., Harri et al. 2014), detailed calibration can yield mixing ratios of water vapor near the surface, which is hitherto the primary indicator of regolith-atmosphere exchange of water (Zent et al. 2016; Savijärvi et al. 2016, 2019). The near-surface water abundance in the morning can be ~3 times greater than that before the sunrise, at both the Phoenix and the Curiosity landing sites, and this rise indicates release of adsorbed water from the regolith (Zent et al. 2016; Savijärvi et al. 2016). Obviously, humidity measurements suffer large uncertainties in the daytime because of the very low relative humidity, despite that the daytime is when the regolith-atmosphere exchange flux maximizes. In addition, reflected solar radiation and thermal emission spectroscopy from orbiters and scattered sky light spectroscopy from landers have been conducted to measure the column abundance of water vapor (e.g., Jakosky & Farmer 1982; Smith 2004; Smith et al. 2006, 2009; Tschimmel et al. 2008; McConnochie et al. 2018). To relate the column abundance to the near-surface abundance, however, will require knowledge of the vertical distribution.

Fig. 1. Schematic illustration of regolith-atmosphere exchange of water on Mars.

Recently, the Tunable Laser Spectrometer (TLS) onboard the Curiosity rover has made the first in-situ measurement of the isotopic composition of water in the Martian atmosphere (Webster et al. 2013). Meanwhile, telescopic observations have shown that the deuterium to hydrogen ratio (D/H) of the atmospheric water column varies from ~8 times to ~4 times the D/H of Earth's oceans (VSMOW[1]) from the northern polar region to the equator (Villanueva et al. 2015). Other telescopic observations found smaller variations but the same trend (Aoki et al. 2015). The Nadir and Occultation for MArs Discovery (NOMAD) onboard the ExoMars Trace Gas Orbiter, currently operating in an orbit of Mars, can measure HDO and constrain D/H in the atmosphere (Vandaele et al. 2018). These recent developments motivate us to consider the isotopic measurement as a diagnostic tool for the regolith-atmosphere exchange.

On Earth, measuring the isotopic composition of water near the surface is a common way to identify the sources of boundary layer water (e.g., Gat 1996; Galewsky et al. 2016). For example, surface water is most depleted in D or $^{18}O$ in the afternoon, and the diurnal variation can be hundreds of per mil (Galewsky 2015). This variation is interpreted as mixing with the free troposphere, which is strongest in the afternoon, and the free tropospheric water vapor being depleted in D. For another example, the isotopic measurements of forest ambient water vapor have allowed partitioning of the water sources into bare soil evaporation and plant transpiration (e.g., Moreira et al. 1997).

The surface environment of Mars is vastly different from that of Earth, in that the surface pressure and temperature are much lower, and the surface temperature has greater diurnal variation. One may therefore expect a greater role of regolith or soil on the D/H on Mars than on Earth. Physical adsorption, insignificant on Earth, comes into play on Mars.

Here we investigate isotopic water exchange between the atmosphere and the regolith and determine its effect on the D/H of the boundary layer water. The goal is to reveal the controlling subsurface processes that affect the diurnal D/H variation, and develop novel stable-isotope diagnoses for regolith-atmosphere exchanges on Mars. The paper is organized as the follows. Section 2 describes the water transport model and the implementation of isotopic fractionation in the model. We also test the model by reproducing the main features of diurnal water variation observed by Curiosity on Mars. We present the results of the D/H variations at varied latitudes and seasons in Section 3, discuss ramification of the findings in Section 4, and conclude in Section 5.

**2. Model**

We construct a one-dimensional model to simulate transport of isotopic water in the Martian regolith and boundary layer. It has a thermal diffusion module, a water transport module, and a boundary layer module. The model includes the isotope fractionation effects of adsorption, condensation, and molecular diffusion. We do not include the effects of subsurface deliquescence (Martín-Torres et al. 2015) in this paper, and will study that subject in the future.

---

[1] Vienna Standard Mean Ocean Water, D/H = $1.5576 \times 10^{-4}$.

We focus on the diurnal variation of the D/H of the boundary layer in this paper, because this variation may provide a direct probe of the amount and the composition of the water vapor flux between the atmosphere and the regolith. On a diurnal timescale, the D/H near the surface is decoupled from the effect of the large-scale circulation, because the latitudinal mixing time for water is a few tens of sols (Montmessin et al. 2005; Villanueva et al. 2015). As such we reduce the regolith-atmosphere exchange problem to a one-dimensional problem concerning the planetary boundary layer and the regolith. Except for the effects of mesoscale transport of water (Steele et al. 2017), our model should provide a good representation of the diurnal variability of the water abundance and isotopic composition in the boundary layer.

**2.1. Thermal Diffusion Model**

We solve the subsurface temperature by the thermal diffusion equation, viz.
$$\frac{\partial T}{\partial t} = D_\text{T} \frac{\partial^2 T}{\partial z^2}, \tag{1}$$
where the thermal diffusivity $D_\text{T}$ is $(I/\rho C)^2$, where $I$ is the thermal inertia and $\rho C$ is the volumetric thermal capacity. We use the typical thermal inertia of the dusty terrains ($I = 150$ J m$^{-2}$ s$^{-1/2}$ K$^{-1}$, Mellon et al. 2000) and the volumetric heat capacity for the Martian aeolian dunes ($\rho C = 10^6$ J m$^{-3}$ K$^{-1}$, Edgett & Christensen 1991). For models of the Gale Crater, we instead use the thermal inertia measured by REMS at Rocknest, a sand patch in Gale Crater ($I = 300$ J m$^{-2}$ s$^{-1/2}$ K$^{-1}$, Martínez et al. 2014).

**2.2 Water Transport Model**

The model includes three components of water in the regolith, i.e.,
$$n_\text{s} = f n_\text{w} + n_\text{d} + n_\text{c}, \tag{2}$$
where $n_\text{s}$ is the total number density of water in the regolith, $n_\text{w}, n_\text{d}, n_\text{c}$ is the number density of water vapor, adsorbed water, and condensed water, respectively, and $f$ is the porosity of the regolith (Zent et al. 1993; Mellon & Jakosky 1993).

Physical adsorption and desorption are assumed to be in equilibrium, and follows a Freundlich isotherm measured for palagonite, a terrestrial analog of the Martian basaltic soil (Zent & Quinn 1995, 1997). The measurements were performed under the Martian conditions, including the temperature and the $CO_2$ partial pressure, and therefore the isotherm inherently includes co-adsorption of $CO_2$. The isotherm takes a form as
$$n_\text{d} m_{\text{H}_2\text{O}} = \rho a_\text{s} \Lambda \left(\frac{K n_\text{w} k_\text{b} T}{1 + K n_\text{w} k_\text{b} T}\right)^\gamma, \tag{3}$$
where $m_{\text{H}_2\text{O}}$ is the mass of a water molecule, $\rho$ is the density of the regolith, $a_\text{s}$ is the specific surface area of the regolith, $\Lambda$ is the mass of water per unit surface area at full and monolayer coverage, $k_\text{b}$ is the Boltzmann constant, $\gamma = 0.4734$ is experimentally measured (Zent & Quinn 1997), and $K$ is
$$K = K_0 e^{\frac{\varepsilon}{T}}, \tag{4}$$

where $K_0 = 7.54 \times 10^{-9}$ Pa$^{-1}$ and $\varepsilon = 2697.2$ K are experimentally measured (Zent & Quinn 1997). We use a typical density for the regolith of $\rho = 1300$ kg m$^{-3}$. $\Lambda$ is scaled from an adsorption of $1.1 \times 10^{-8}$ kg m$^{-2}$ at $T = 209$ K and a water partial pressure of 0.34 Pa (Zent & Quinn 1997). We adopt a nominal value for the specific area of 100 m$^2$ g$^{-1}$ and explore the impact of a lower value of 17 m$^2$ g$^{-1}$. This range is based on in-situ measurements of Viking (Ballou & Wood 1978) and measurements of the JSC Mars-1 analog (Meslin et al. 2011).

We assume condensation is in equilibrium. In other words,
$$n_c = \max[0, n_s - f n_{w,sat} - n_d(n_{w,sat}, T)], \tag{5}$$
where $n_{w,sat}$ is the saturation vapor pressure of water. This equation implies that when condensation occurs, the vapor pressure in the pore space is kept at the saturation vapor pressure.

The diffusion equation is (e.g., Mellon & Jakosky 1993)
$$\frac{\partial n_s}{\partial t} = \frac{\partial}{\partial z}\left(D \frac{f}{\tau} \frac{\partial n_w}{\partial z}\right), \tag{6}$$
where $D$ is the diffusivity of water and $\tau$ is the tortuosity of the regolith. We typically assume a soil porosity of 0.5, and a tortuosity of 1.5 (Mellon & Jakosky 1993; Sizemore & Mellon 2008; Meslin et al. 2011). Using Equation (2), one can derive
$$\frac{\partial n_w}{\partial z} = \frac{\frac{\partial n_s}{\partial z} - \frac{\partial n_d}{\partial T}\frac{\partial T}{\partial z} - \frac{\partial n_c}{\partial z}}{f + \frac{\partial n_d}{\partial n_w}}. \tag{7}$$
This equation has the same form as Eq. 13 of Zent et al. (1993). The denominator of the above equation indicates that one effect of adsorption is to retard diffusion, typically by several orders of magnitude (e.g., Jakosky 1985; Mellon & Jakosky 1993; Zent & Quinn 1997). $D$ is a function of the temperature and the pore size of the regolith, and is calculated by combining the molecular diffusion coefficient ($D_{12}$) and the Knudsen diffusion coefficient ($D_{1K}$), in the same way as Mellon & Jakosky (1993). We use
$$D_{12} = C_1 \left(\frac{T^{3/2}}{P\Omega_{12}}\right), \tag{8}$$
and
$$D_{1k} = C_2 T^{1/2} r, \tag{9}$$
where $P$ is the pressure in Pascal, $\Omega_{12}$ is the collision integral (Eq. 12 of Mellon & Jakosky 1993), $r$ is the pore size in cm, the constant $C_1$ is 4.865 for H$_2$O, 4.774 for HDO, and 4.690 for H$_2^{18}$O, and the constant $C_2$ is 2286 for H$_2$O, 2226 for HDO, and 2169 for H$_2^{18}$O for $D_{12}$ and $D_{1K}$ having a unit of cm$^2$ s$^{-1}$. When the pore is small, Knudsen diffusion dominates, and when the pore is large, molecular diffusion dominates. The transition occurs at 6 μm under the Martian conditions. The nominal model uses a pore size of 20 μm, consistent with the typical thermal inertia (Presley & Christensen 1997; Piqueux & Christensen 2009). This pore size makes the diffusion safely in the molecular diffusion regime, and renders the results insensitive to the exact value used for the pore size. We additionally explore a pore size of 1 μm, where Knudsen diffusion dominates.

We do not include in this work the reduction of the porosity and the pore size if condensation of water occurs in the regolith. Since the diurnal exchange of water is limited to the top centimeter of the regolith, and we only apply the model to locations and seasons outside the seasonal polar caps, ice is not expected to fill the pore (Mellon & Jakosky 1993). We later verify that the ice accumulates to up to 0.001 m³/m³ in our model scenarios, justifying this assumption.

For atmospheric water we apply a simplified model without condensation. The abundance then follows a passive tracer equation, i.e.,

$$\frac{\partial n_a}{\partial t} = \frac{\partial}{\partial z}\left(k_{zz} N_a \frac{\partial (n_a/N_a)}{\partial z}\right), \tag{10}$$

where $n_a$ is the number density of water vapor in the atmosphere, $N_a$ is the total number density of the atmosphere, and $k_{zz}$ is the eddy diffusion coefficient.

Finally, the flux from the regolith to the atmosphere is

$$\Phi = v_{\text{dep}}(n_w - n_a), \tag{11}$$

where $v_{\text{dep}}$ is the deposition velocity, and $n_w$ and $n_a$ are values taken at the surface. In a microscopic scale, the mass transfer through the regolith-atmosphere boundary takes two steps. One is molecular diffusion through a quasi-laminar layer of thickness $z_0$, and the other is turbulent transport from the lowest atmospheric layer to a distance of $z_0$ from the surface. The deposition velocity is then the combination of the aerodynamic resistance ($r_a$) and the quasi-laminar resistance ($r_b$), viz.

$$v_{\text{dep}}^{-1} = r_a + r_b. \tag{12}$$

The aerodynamic resistance needs to be calculated by a boundary layer model, and the quasi-laminar resistance is (Seinfeld & Pandis 2006)

$$r_b = \frac{5 \text{Sc}^{2/3}}{u_*}, \tag{13}$$

where Schmidt number $\text{Sc} = \nu/D_{12}$, $\nu$ is the kinematic viscosity of the atmosphere, and $u_*$ is the friction velocity which will be defined in Section 2.3. In practice we find that $r_a$ is greater than $r_b$ by approximately three orders of magnitude. In other words, it is the aerodynamic resistance that controls the exchange flux. Because the aerodynamic resistance does not depend on the mass of molecule (see below), the deposition velocity has little mass dependency and does not cause fractionation.

**2.3 Boundary Layer Model**

The water transport model calls for the eddy diffusion coefficient ($k_{zz}$) and the deposition velocity ($v_{\text{dep}}$). A boundary layer model provides estimates of these two parameters. Classical work in modeling the planetary boundary layer of Mars and its coupling with the regolith includes Sutton et al. (1979), Haberle et al. (1993), Zent et al. (1993), and Savijärvi (1995, 2012). The boundary layer models of Haberle et al. (1993) and Savijärvi (1995, 2012) solve the same momentum and heat equations in slightly different ways. Key features of the Martian boundary layer include (1) setup of the boundary layer after dawn and subsequent extension as the surface temperature rises, (2) collapse of the boundary layer in the late afternoon when the surface temperature becomes lower than the ambient temperature, and (3) a vertically stratified and stable atmosphere during the night (Fig. 1).

In this work, we use the bulk aerodynamic method developed for Mars applications by Sutton et al. (1979). The method has three steps: (1) determine a bulk Richardson number from the wind speed and the difference between the atmospheric temperature and the ground temperature, (2) determine the Monin-Obukhov length from the bulk Richardson number and the surface roughness length, (3) estimate the drag and the heat transport coefficients from the Monin-Obukhov length. The input parameters are the wind speed, the difference between the atmospheric temperature and the ground temperature, and the surface roughness length ($z_0$).

The bulk Richardson number is
$$\text{Ri}_\text{B} = \frac{gz_1 \Delta\theta}{u^2 \bar{T}}, \quad (14)$$
where $g$ is the surface gravity, $z_1$ is the height at which the wind speed is defined or measured (e.g., 1.6 m for REMS), $\Delta\theta$ is the atmosphere-surface temperature difference (positive when the atmosphere is hotter), $u$ is the wind speed, and $\bar{T}$ is the mean between the ground and the atmosphere temperatures. The bulk Richardson number is linked to the Monin-Obukhov length ($L$) by $\text{Ri}_\text{B} = f(z_1, z_0, L)$, where $f$ is an integral function from the boundary layer similarity theory and empirical terrestrial measurements (Eqs. 5-6 of Sutton et al. (1979)) and $z_0$ is the surface roughness length. The surface roughness length is uncertain, and we adopt a nominal value of 0.1 cm and explore the impact of a larger value of 1 cm following Sutton et al. (1979). This equation always has a single solution for $L$, except for highly stable cases when the bulk Richardson number exceeds a critical value for the extinction of turbulence ($\text{Ri}_\text{B} \sim 0.2$). This situation rarely occurs at the surface – when it does, a small value can be assigned to the mass transfer coefficients.

We can then estimate the parameters relevant for regolith-atmosphere exchange. The friction velocity is $u_* = u C_d(z_1, z_0, L)$, where $C_d$, the drag coefficient, is another known integral function (Eq. 9 of Sutton et al. (1979)). The heat diffusivity $K_H$ is
$$K_H = \frac{\kappa u_* z}{\phi_H(z/L)}, \quad (15)$$
where $\kappa \sim 0.4$ is the von Karman constant, and $\phi_H$ is the empirically determined temperature profile function (Eq. 6 of Sutton et al. (1979)). The equation above is valid between $z_0$ and $z_1$. It is generally assumed that $K_{zz} \sim K_H$, i.e., the turbulence is responsible to both the mass and the heat vertical transport (e.g., Zent et al. 1993). The aerodynamic resistance $r_\text{a}$ is then
$$r_\text{a} = \int_{z_0}^{z_1} \frac{\phi_H(z/L)}{\kappa u_* z} dz. \quad (16)$$

The bulk aerodynamic method only estimates the value of $K_{zz}$ near the surface ($z_1$). We however need the value of $K_{zz}$ in the entire atmospheric column for completeness. For simplicity we use the expression of $K_H$ applies to $z > z_1$, while taking $\phi_H \sim \phi_H(z_1/L)$, and apply the Blackadar's mixing length formulation (Blackadar 1962)
$$K_{zz} = \frac{K_H}{1 + K_H/K_\infty}, \quad (17)$$
where $K_\infty$ is the eddy diffusion coefficient of the free troposphere which we adopt the typical value of 100 m² s⁻¹. With Eq. 17, $K_{zz}$ asymptotes to $K_\infty$ when $z$ is large. We will show later in Section 2.6 that this simplified approach produces results that are consistent with the diurnal variation of the water abundance measured in Gale Crater.

In summary, our water transport model for the regolith and the boundary layer takes the basic soil properties (i.e., thermal inertia, volumetric heat capacity, pore size, porosity, tortuosity, specific surface area, and surface roughness length), as well as the meteorological data (the wind speed and the near-surface and the ground temperatures) as input parameters. The latter can be measured on Mars or calculated by the general circulation models.

**2.4 Isotopic Fractionation**

Regolith adsorption fractionates water because the latent heat of adsorption of HDO is higher than that of $H_2O$. Moores et al. (2011) conducted experiments of water transport in JSC Mars-1, a commonly used Martian regolith analog, and measured the effective diffusivities of $H_2O$ and HDO under temperatures, pressures, and the background atmosphere corresponding to Mars. The experiment found a smaller effective diffusivity of HDO than $H_2O$, implying that more HDO is partitioned in the adsorbed phase than $H_2O$ at the same condition. According to Equation (7), the ratio between the effective diffusivity is the inversed ratio between the adsorption coefficient of HDO and $H_2O$. As the difference in the latent heat drives the difference in the adsorption coefficient, the fractionation factor should follow an Arrhenius form at low temperatures, i.e., $\alpha_d \propto e^{\Delta H/RT}$, where $\Delta H$ is the difference of the latent heat of adsorption, and $R$ is the gas constant (Criss 1999). We fit the effective diffusivity data of Moores et al. (2011) to the form and obtain the following fractionation factor due to adsorption

$$\alpha_d \equiv \frac{n_d^*/n_d}{n_w^*/n_w} \sim \frac{D_{H_2O, \text{EFF}}}{D_{HDO, \text{EFF}}} = e^{\max\left(0, \frac{902.84 \pm 433.54}{T} - 3.6255 \pm 1.9573\right)}, \quad (18)$$

where $T$ is in K, and the quantities with the superscript * denote the quantities for HDO. The 1-σ standard deviation of the fitted parameters are provided. This formula implies $\Delta H \sim 7 \pm 3 \text{ kJ mol}^{-1}$. There are no experimental data points for $T>240$ K, but the fractionation factor approaches unity at the high-temperature end of the experiments. The quality of the existing data does not allow us to assess whether a crossover exists or a $1/T^2$ term is needed. We thus assume the fractionation factor to be unity at temperatures higher than ~240 K.

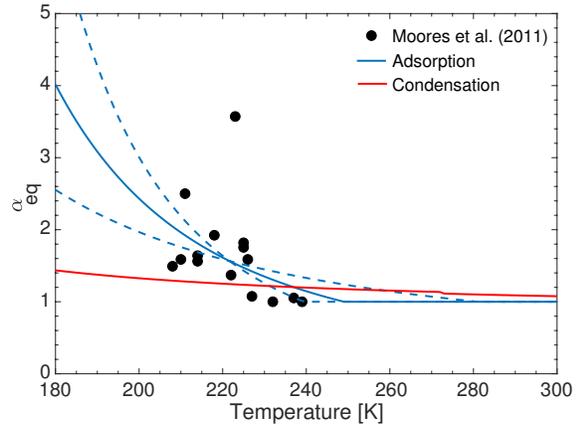

Fig. 2. Isotopic fractionation factors used in the model. The fractionation factor of adsorption (in blue) is obtained by fitting to the measurements of Moores et al. (2011). The dashed curves show the fractionation factor from the 1-σ standard deviation of the fitted parameters.

Fractionation by condensation is caused by the vapor pressure isotope effect, as

$$\alpha_c \equiv \frac{n_c^*/n_c}{n_w^*/n_w} \sim \frac{n_{\text{sat}, H_2O}}{n_{\text{sat}, HDO}}. \quad (19)$$

We use the fractionation factor between gas and liquid from Jancso & Van Hook (1974) when $T>273.15$ K and that between gas and solid from Lamb et al. (2017) for $T<273.15$ K. The formula for $\alpha_c$ for $T<273.15$ K is

$$\alpha_c = e^{\left(\frac{13525}{T^2} - 0.0559\right)}. \tag{20}$$

As shown in Fig. 2, it is likely that adsorption is more effective in fractionating water than condensation under the conditions of a typical Martian night. We caution that the scatter of the experimental data is rather large, and there are no data for temperatures less than 205 K. The fitted Arrhenius formula may thus overestimate the isotope effect of adsorption at very low temperatures.

Finally, we include the diffusion isotope effect by applying different diffusivities for HDO and H$_2$O. For HDO, we have

$$n_s^* = f n_w^* + n_d^* + n_c^*. \tag{21}$$

For equilibrium fractionation,

$$n_d^* = n_w^* \left(\frac{n_d}{n_w} \alpha_d\right) \equiv n_w^* A, \tag{22}$$

$$n_c^* = n_w^* \left(\frac{n_c}{n_w} \alpha_c\right) \equiv n_w^* B. \tag{23}$$

As such, Equation (21) can be rewritten as $n_s^* = (f + A + B) n_w^*$. The diffusion equation for HDO is

$$\frac{\partial n_s^*}{\partial t} = \frac{\partial}{\partial z}\left(D^* \frac{f}{\tau} \frac{\partial n_w^*}{\partial z}\right), \tag{24}$$

where $D^*$ is the diffusivity of HDO, and

$$\frac{\partial n_w^*}{\partial z} = \frac{\frac{\partial n_s^*}{\partial z} - n_w^* \frac{\partial A}{\partial z} - n_w^* \frac{\partial B}{\partial z}}{f + A + B}. \tag{25}$$

This way of combining of the equilibrium isotope effects and the diffusion isotope effect has been widely used in the models of evaporation from unsaturated soil (e.g., Barnes & Allison 1984; Mathieu & Bariac 1996), but the formulation presented here additionally traces the adsorbed water and includes the fractionation due to adsorption. The assumption of the equilibrium fractionation is the natural consequence of a broader assumption of equilibrium in both condensation and adsorption. In our model, the saturation ratio is always unity when condensation occurs, and this effectively eliminates any kinetic isotope effect in condensation (e.g., Casado et al. 2016). Deviation from this assumption will require treating condensation and adsorption as kinetic processes, and is beyond the scope of this work.

**2.5 Numerical Procedure**

The thermal diffusion and the subsurface water transport equations are solved by an implicit Euler method on a vertical grid from zero to 10 thermal skin depths (typically 6 m for a diurnal variation). It is reasonable to cut the lower boundary solely based on the thermal skin depth because the effective diffusivity of water vapor is typically several orders of magnitude smaller than the thermal diffusivity due to adsorption (e.g., Mellon & Jakosky 1993; Zent et al. 1993). The thickness of the first sub-surface layer is 1 mm, and each subsequent layer is 50% thicker.

The transport of water in the atmosphere is solved on a vertical grid from zero to 50 km, well above the top of the planetary boundary layer at any time (e.g., Haberle et al. 1993). The thickness of the first atmospheric layer is 1 m, and each subsequent layer is 50% thicker.

All number densities ($n_a$, $n_s$, $n_w$, $n_d$, $n_c$) are defined at the center of each atmospheric and regolith layers. As such, the mass flux between adjacent layers is naturally defined. The numerical scheme is therefore backward difference in time and central difference in space, and is strictly mass conserving.

The first step is to calculate the subsurface temperature. This step is separable from the water transport calculation because the heat deposit by phase transition of water is negligible. We have verified that the heat flux produced by water adsorption and condensation in the regolith is at least three orders of magnitude less than the heat flux of conduction. In other words, the subsurface temperature is entirely driven by thermal conduction. We start from an isothermal temperature profile. The upper boundary condition is specified by either measurements or models of the surface temperature. We apply a zero-flux lower boundary condition, because the geothermal flux is negligible for the vertical scale of this problem. The time-stepping integration continues to the point when the maximum difference between temperatures from successive periods is less than 0.1%.

The second step is to calculate the water concentration in the atmosphere and the regolith. We apply a zero-flux boundary condition at both the upper boundary of the atmosphere and the lower boundary of the regolith. In other words, we simulate a closed system where water can only redistribute between the atmosphere and the regolith. $n_a$ and $n_s$ are the independent parameters in the time stepping, and $n_s$ is partitioned into $n_w$, $n_d$, and $n_c$ after each step. We start with a specified initial column mass of water, and the total mass of water in the regolith-atmosphere system is kept constant in the model run. The time-stepping integration continues to the point when the maximum difference between number densities from successive periods is less than 0.1%. We use a time step of 1/20 of a Martian hour to resolve the diurnal variation.

**2.6 Testing the Model for Diurnal Water Exchange**

We test our water transport model and compare model results with the relative humidity measurements made in situ by Curiosity's REMS instrument. In particular, we obtain the REMS re-calibrated data of relative humidity, air temperature, surface temperature, and wind speed from the Planetary Data System (PDS). We only use the data taken during the first 10 s after the relative humidity sensor has been turned on, because the sensor heats up when making measurements (Rivera-Valentín et al. 2018). We choose two seasons for this test. One is $L_s \sim 195°$, or Sol 78, when the atmospheric water column is in the highest range due to sublimation of the north polar cap, and the other is $L_s \sim 71°$ (aphelion) or Sol 501, when the relative humidity is in the highest range.

Besides the soil properties, for which we adopt the nominal values in this test, the only free parameter is the initial amount of water in the model column. We pick an initial amount of water so that the resulting water column in the atmosphere matches the value measured by Curiosity's Chemistry & Camera (ChemCam) instrument in each season (McConnochie et al. 2018). We do not fine tune the model to create a fit; rather, we intend to test whether the model gives sensible results in terms of the distribution of water between the regolith and the atmosphere and its diurnal variation.

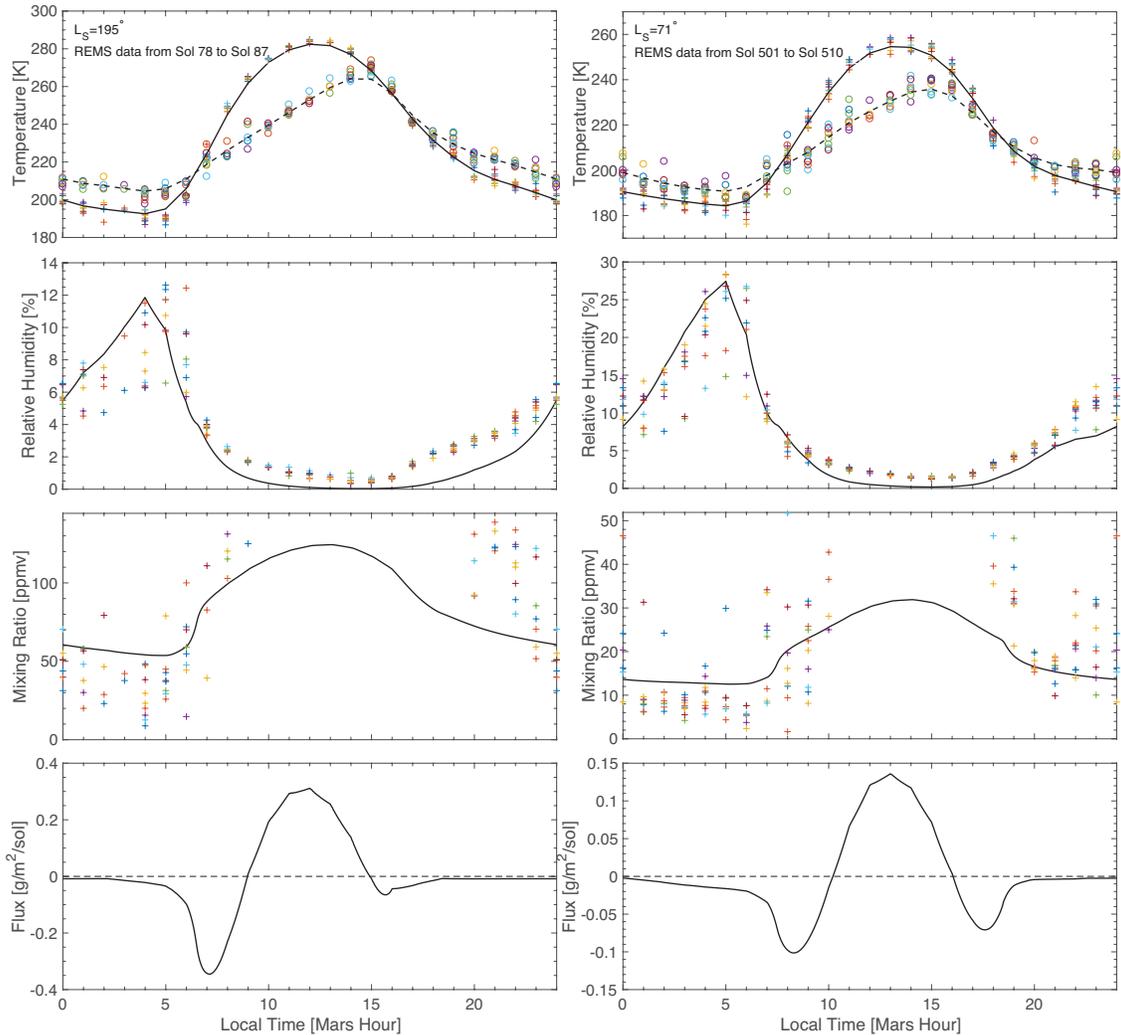

Fig. 3. Modeled diurnal variation of humidity in Gale Crater in comparison with the REMS data. The first row shows the averaged REMS measurements of the surface temperature (solid line) and the ambient temperature (dashed line) in the 10-Sol period for each season as input parameters. The second to fourth rows show the relative humidity, the absolute humidity, and the regolith-atmosphere exchange flux, respectively. A positive flux means regolith emits water into the atmosphere. The predicted humidity variation is generally consistent with the REMS data.

Fig. 3 shows the model results in comparison with the measured relative humidity and absolute humidity (expressed as the water vapor mixing ratio). Most of the variation seen in the relative humidity is caused by the variation of the temperature; however, the measured absolute humidity still features a rise by a factor of 2 – 3 after sunrise. This feature is well captured by the model. Note that the absolute humidity around midday cannot be measured because the relative humidity is extremely low. This is a bona fide test because for a fixed water column, should there be major errors in the model, the nighttime absolute humidity would depart substantially from the data. The diurnal trend of the surface water abundance predicted by our model is also consistent with the existing 1D model (Savijärvi et al. 2019) and mesoscale 3D model (Steele et al. 2017) for both seasons. A common feature in REMS data that neither of these 1D and 3D models, including the one described here, could reproduce is the somewhat elevated humidity in the evening before midnight. It is unclear whether additional processes may cause this elevated humidity.

The model results in Gale Crater show key diurnal patterns of regolith-atmosphere exchange flux of water on Mars, first described in Zent et al. (1993) and later shown repeatedly by models (e.g., Savijärvi et al. 2016). In Fig. 3 we can see (1) a small flux from the atmosphere to the regolith during the night, (2) an enhanced flux from the atmosphere to the regolith in a short period around dawn, due to the setup of the boundary layer and the ground temperature being still low, (3) a flux from the regolith to the atmosphere during around midday, (4) a moderate flux from the atmosphere to the regolith in the afternoon when the ground temperature drops, and (5) the flux decreases in the evening as the boundary layer collapses. In sum, our water transport model captures the key processes of regolith-atmosphere exchange on Mars and produces results in good agreement with both data and previous models.

## 3 Results

The modeled D/H diurnal variation in Gale Crater is shown in Fig. 4. The starting atmospheric D/H is chosen such that the resulting δD matches the measurements made by TLS on Curiosity (δD=4950±1080‰ with samples taken at ~23H, Webster et al. 2013). The D/H is driven by the fractionation in adsorption, and no condensation is found in these models. Since the fractionation factor is larger at a lower temperature, the magnitude of the variation is larger for a colder season. The D/H variation is ~900‰ at the surface and ~200‰ at 100 m during the season of aphelion. The magnitude of the D/H variation is sensitive to the fractionation factor of adsorption. The 1-σ uncertainty of the temperature dependence of the fractionation factor

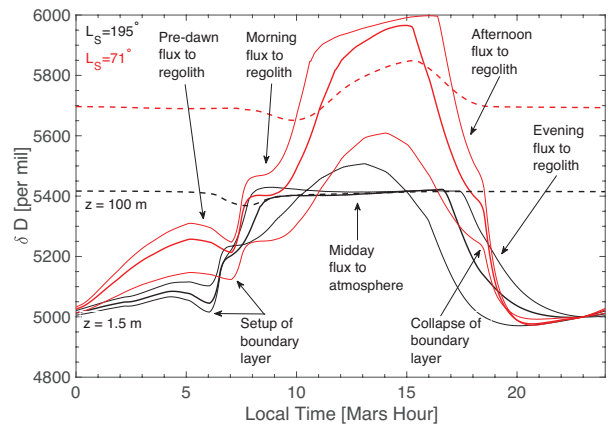

Fig. 4. Modeled diurnal variations of D/H in Gale Crater. Thick lines are the nominal model of adsorption fractionation, and thin lines are produced using the standard deviations of fractionation factor. The variation at the surface is up to ~900‰ at the season of aphelion.

results in the magnitude of the diurnal variation ranging between 600 and 1000‰ at the surface.

The D/H variation at the surface reveals a complex interaction between the regolith-atmosphere exchange and atmospheric mixing. The main effects seen in the D/H variation are summarize as the follows. (1) Shortly before the sunrise, the small flux from the atmosphere to the regolith causes δD at the surface to decrease. (2) Right after the sunrise and the setup of the boundary layer, mixing of the boundary layer causes δD of the bottom atmosphere to rise, while δD at 100 m remains unchanged. (3) Shortly after, the enhanced flux from the atmosphere to the regolith causes δD to decrease again – this time the effect is seem at both the surface and at 100 m. (4) In the late morning and around midday, the flux from the regolith to the atmosphere causes δD to rise, and it maximizes in the early afternoon. The fractionation stops when the surface temperature becomes >240 K (Fig. 2). The regolith-atmosphere exchange is rapid in this period due to the turbulent boundary layer and the rapid change of the surface temperature, and the variation of δD of the bottom atmosphere is driven by Rayleigh fractionation. (5) In the late afternoon the flux from the atmosphere to the regolith causes δD at both the surface and at 100 m to decrease quite rapidly. Similar to the previous period, the δD variation in this period can be approximately by Rayleigh fractionation. (6) In the evening after the collapse of the boundary layer, δD at the surface first decreases rapidly due to adsorption in the regolith and then rise slowly due to slow atmospheric mixing, while δD at 100 m remains unchanged. Here we see that other than the periods (5) and (6), the δD variation is driven by temperature-dependent adsorption equilibrium coupled with diffusion in the regolith and mixing in the atmospheric boundary layer.

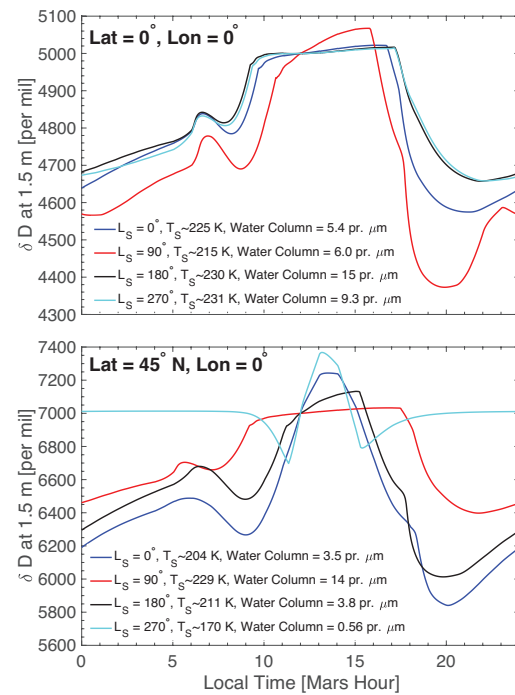

Fig. 5. Modeled diurnal D/H variations at the equatorial and mid-latitude locations of Mars. Labels show the mean surface temperature and the atmospheric water column, which we make to match with orbital remote sensing (Smith et al. 2009) by adjusting the initial water loading. The magnitude of the D/H variation is greater for a lower surface temperature, except when condensation occurs.

To study more generally the diurnal variation of D/H at various latitudes and seasons, we choose three representative latitudes: equatorial (Lat~0°), mid-latitude (Lat~45°), and polar (Lat~68°), and use the Mars General Circulation Model (MGCM) outputs (Forget et al. 1999) made available via the Mars Climate Database (http://www-mars.lmd.jussieu.fr/mars/access.html) for the surface temperatures and the wind speed. The MGCM contains a sophisticated treatment for the boundary layer (Colaïtis et al. 2013). We explore four seasons for the equatorial and mid-latitude locations, and focus on the seasons that are well outside of the seasonal polar cap for the polar location. For Lat~68°, the surface is free of $CO_2$ ice deposits approximately from $L_s$~30° to $L_s$~210° (Kieffer & Titus 2001). We choose the starting atmospheric D/H such that the resulting δD at noon is 5000‰ at the equatorial and 7000‰ at the mid-latitude and polar locations (Villanueva et al. 2015). The δD variation is scalable with respect to its mean value because the transport equation is linear with respect to the abundance of HDO. The results are shown in Figs. 5 and 6.

The magnitude of the diurnal variation of D/H is on the order of 400 – 800‰ in the equatorial location and 600 – 1400‰ in the mid-latitude location. The magnitude of the diurnal variation changes with season, and it becomes greater during a colder season. This is because the fractionation factor is greater when the temperature is colder (Fig. 2). This trend is valid until condensation occurs. In our sample, condensation in the regolith occurs at the latitude of 45° and at the season of $L_s$~270° (cyan line in the right panel of Fig. 5), and at the latitude of 68° (Fig. 6). We see that the magnitude of the D/H variation is only 600‰ at the latitude of 45° and at the season of $L_s$~270°, even though the mean surface temperature is only 170 K. This is because when condensation occurs, the dominating water phase in the regolith is the condensed phase, and the fractionation factor of condensation is smaller than that of adsorption (Fig. 2). The case at the latitude of 68° and at the season of $L_s$~120° (red line in Fig. 6) is special: condensation also occurs before sunrise (i.e., the coldest time of day) at the surface (i.e., surface frost formation). This surface condensation causes a small dip in the D/H curve.

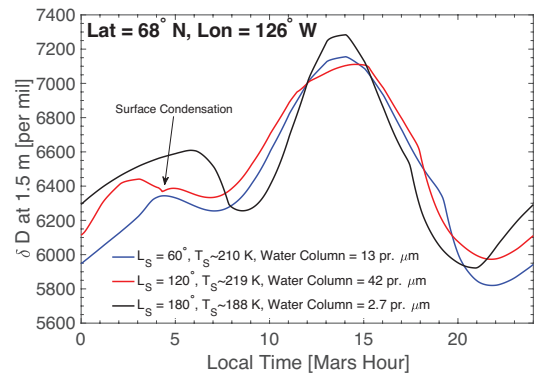

The shape of the diurnal variation is similar when the driving forces for the variation are regolith adsorption and desorption and their coupling with the boundary layer. The adsorbed water is enriched in HDO compared to the water vapor. As a result, when there is a flux from the regolith to the atmosphere, we see an increase of the near-surface D/H (i.e., the midday rise), and when there is a flux from the atmosphere to the regolith, we see a decrease of the near-surface D/H (i.e., the morning and evening drop). In the evening when regolith-atmosphere exchange is weak, the near-surface D/H gradually returns to the diurnal average due to mixing in the atmosphere.

Fig. 6. Modeled diurnal D/H variations at a polar location in the province of the Phoenix landing site. The $L_S$=120° model has surface condensation that causes a small dip in the D/H curve.

The shape becomes different when condensation occurs, mainly in that the rise of D/H is delayed (see cyan line in the right panel of Fig. 5). This is because when the temperature of the regolith rises, water in the regolith experiences a phase change from the condensed phase to the adsorbed phase. What is the isotopic effect of this phase change? Because the adsorbed phase is more enriched in D than the condensed phase, this phase change removes D from the gas phase, and therefore it causes the atmospheric D/H to drop even when the regolith emits water into the atmosphere. The effect is eventually overcome by the mass release of adsorbed water to the atmosphere, when the temperature becomes sufficiently high, and thus causing the delayed rise of D/H. As such, the strength and the timing of the D/H variation provide a measure of the flux of regolith-atmosphere exchange and the phase transition of water in the regolith.

The diurnal variations of D/H described here depend on the fractionation factors being used (Fig. 7, right panel). The extrapolation of the fractionation factor to temperatures less than 200 K would cause the resulting D/H prediction to be uncertain by approximately 30%. The D/H variation is however insensitive to the water abundance to the extent that no condensation occurs, because the adsorption occurs at the linear portion of the adsorption isotherm, and all number densities are proportional to the overall water abundance (Fig. 7, left panel).

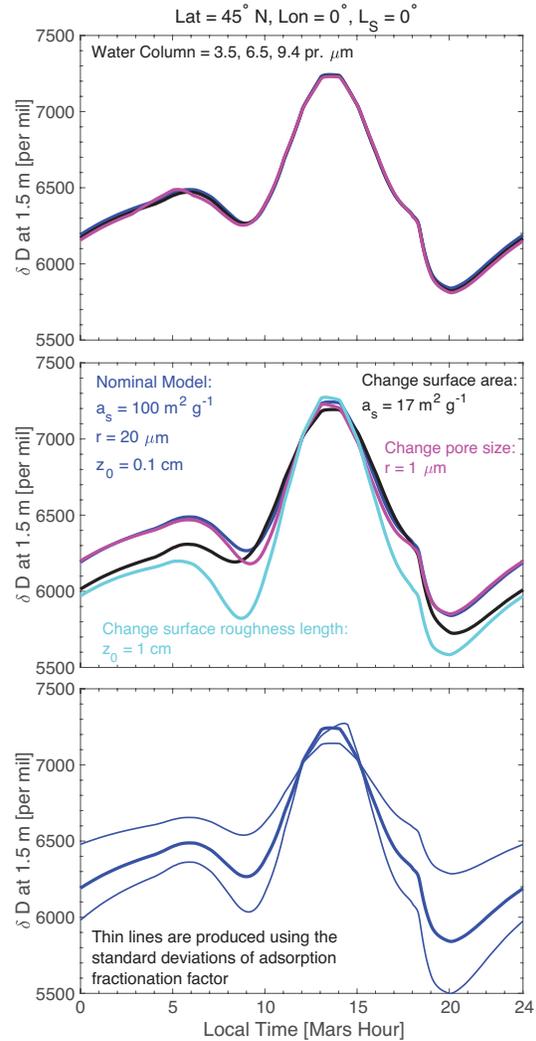

Fig. 7. Sensitivity of the diurnal D/H variation to the atmospheric water column (top), the regolith's surface area, pore size, surface roughness (center), and the fractionation factor of adsorption (bottom).

Changing the surface area or the pore size in the Mars-relevant range has little effects either (Fig. 7, center panel). Changing the surface roughness length to 1 cm would double the deposition velocity, increase the exchange flux by ~60%, and increase the magnitude of the diurnal D/H variation from 1400‰ to 1650‰ at the latitude of 45° and at the season of $L_s$~0° (Fig. 7). The sensitivity of the boundary layer quantities to the surface roughness length has also been identified, and is much reduced when a molecular thermal sublayer is additionally considered (Sutton et al. 1979).

**4. Discussion**

Measuring the diurnal variation of D/H in water on Mars's surface with sufficient precision would be a new indicator for the regolith-atmosphere exchange of water. The examples shown in this paper indicate that the D/H variation traces the exchange flux. The variation is principally driven by the adsorption-desorption cycle as the temperatures of the ground and the regolith change. The D/H variation at the surface is also affected by the setup and collapse of the boundary layer and mixing in the atmosphere. This variation would appear to be well separated from the variation caused by surface frost formation or precipitation, which should follow Rayleigh fractionation (Fig. 8).

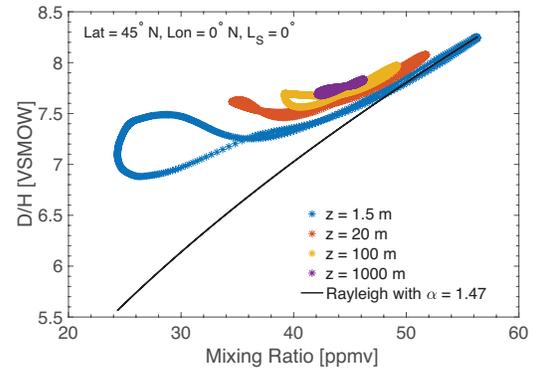

Fig. 8. D/H versus water mixing ratio, in comparison with Rayleigh fractionation at the mean surface temperature of 205 K.

In addition to physical adsorption, the D/H variation provides a window into the phase transition inside the regolith. We have shown that condensation in the regolith and at the surface causes distinctive features in the D/H curve (Figs. 5 and 6). Another potential phase transition is deliquescence – formation of liquid brines of perchlorate at relative humidities well less than 100% (Martín-Torres et al., 2015; Zent et al. 2016; Rivera-Valentín et al. 2018). The effect of deliquescence on the D/H variation warrants further studies.

What precision do we need to detect the D/H variation predicted in this paper? At the surface, the magnitude of the variation is on the order of ~1000‰ depending on the specifics of the local climate and soil properties. The TLS instrument onboard Curiosity has reported a D/H precision of 1000‰. An improvement of the precision by a factor of ~5 and a sampling rate on the order of hours would allow detection of the predicted D/H variation. Since the current TLS instrument determines the water abundance via weak lines in a $CO_2$ band (Webster et al. 2013), the desired precision may well be achieved if the instrument is instead optimized for a $H_2O$ band.

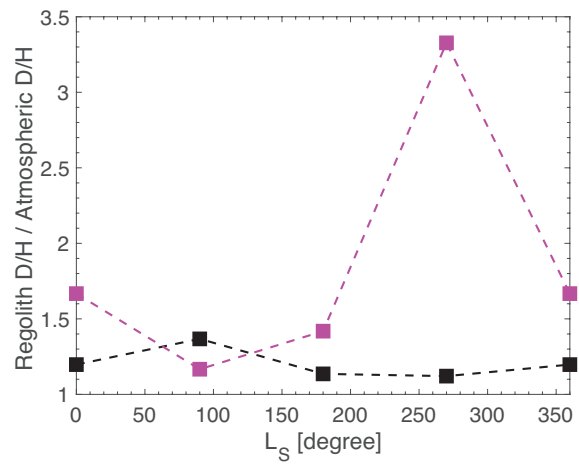

Fig. 9. Diurnal mean D/H of the top 10-cm regolith water (adsorbed or condensed) with respect to the D/H of the atmospheric column water, for Lat~0° (black) and Lat~45° (purple), respectively. The regolith water is expected to exchange with the atmospheric column water on a seasonal timescale.

Although the diurnal exchange of water typically extends to a depth of only 1 cm, seasonal exchange extends to ~10 cm. Our model indicates that the water adsorbed or condensed in this deeper part of the regolith has much greater D/H compared to the atmosphere, and is changing from season to season (Fig. 9). This water is exchangeable with the atmospheric column seasonally. In other words, the regolith is expected to substantially affect the D/H in the atmospheric water column. The observed distribution of D/H from the north pole to the equator (Villaneuva et al. 2015) appears to be steeper than what is predicted by a GCM (Montmessin et al. 2005). The regolith may have a role in shaping the distribution and causing seasonal variations, and its effect may be observed by the NOMAD instrument onboard the ExoMars Trace Gas Orbiter via solar occultation (e.g., Vandaele et al. 2018).

**5. Conclusions**

We have developed a regolith-atmosphere exchange model and applied to the model to study the exchange of isotope water on Mars. The model reproduces the diurnal changes of the near-surface water vapor abundance observed by Curiosity in Gale Crater. Coupled with the equilibrium fractionation factors between HDO and $H_2O$ in physical adsorption and condensation, the model predicts substantial variation of D/H in near-surface water vapor. This variation ranges from 300‰ to 1400‰, and is greater at higher latitudes or colder seasons. This variation is principally driven by physical adsorption to the regolith, and its change as the temperature of the regolith changes. At high latitudes and during winter, condensation occurs in regolith and alters the diurnal pattern of the D/H variation.

The predicted magnitude of the D/H variation depends on the fractionation factor in adsorption and its temperature dependency. This research thus highlights the need to further quantify the fractionation factor by experiments. The predicted D/H variation can be tested by in-situ experiments in the future. The isotopic measurements, complementary with the humidity measurements, will pinpoint the water exchange flux between the regolith and the atmosphere in both the daytime and the nighttime, and allow attribution of this flux to specific processes in the regolith.

**Acknowledgement**

I thank Yuk Yung and Bethany Ehlmann for helpful discussion on Mars's water cycle, and Bruce Jakosky and another reviewer for careful review of the manuscript. The research was carried out at the Jet Propulsion Laboratory, California Institute of Technology, under a contract with the National Aeronautics and Space Administration. RH was supported by NASA's Habitable Worlds grant #NNN13D466T.18